\documentclass{actasen}

\usepackage{diagbox,multirow}
\begin{document}

\setcounter{page}{101}

\Volume{2018}{42}


\runheading{SUN Zhao et al.}

\title{Photometric Follow-up Transit (Primary Eclipse) Observations of WASP-43 b and TrES-3b and A Study on Their Transit Timing Variations$^{\dag}~ \!^{\star}$}


\footnotetext{$^{\dag}$ Supported by National Natural Science
Foundation (11273068, 11473073, 11303102, 11573073), Special Project of Strategic and Guiding Science and Technology of the Chinese Academy of Sciences (Type B)(XDB09000000), Project for the Layout of Developing and Interdisciplinary Sciences of the Chinese Academy of Sciences (KJZD-EW-Z001), and Asteroid Foundation of Purple Mountain Observatory

\hspace*{3mm}Received 2016--04--27; revised version 2016--05--16

$^{\star}$ A translation of {\it Acta Astron. Sin.~}
Vol.57, No.6, pp. 696--718, 2016 \\
\hspace*{5mm}$^{\bigtriangleup}$ sun\_zhao\_0@hotmail.com\\
\hspace*{5mm}$^{\bigtriangleup}$$^{\bigtriangleup}$jijh@pmo.ac.cn\\

\noindent 0275-1062/01/\$-see front matter $\copyright$ 2018 Elsevier
Science B. V. All rights reserved. 

\noindent PII: }


\enauthor{SUN Zhao$^{1,2}$$^{\bigtriangleup}$ \hs\hs JI Jiang-hui$^{1,3}$$^{\bigtriangleup}$$^{\bigtriangleup}$\hs\hs DONG Yao$^{1,3}$}
{\up{1}Purple Mountain Observatory, Chinese Academy of Sciences, Nanjing 210008\\
\up{2}University of Chinese Academy of Sciences, Beijing 100049\\
\up{3}Key Laboratory of Planetary Sciences, Chinese Academy of Sciences, Nanjing 210008\\}


\abstract{Two photometric follow-up transit (primary eclipse) observations on WASP-43 b and four observations on TrES-3 b are performed using the Xuyi Near-Earth Object Survey Telescope. After differential photometry and light curve analysis, the physical parameters of the two systems are obtained and are in good match with the literature. Combining with transit data from a lot of literature, the residuals ($O-C$) of transit observations of both systems are fitted with the linear and quadratic functions. With the linear fitting, the periods and transit timing variations (TTVs) of the planets are obtained, and no obvious periodic TTV signal is found in both systems after an analysis. The maximum mass of a perturbing planet located at the 1:2 mean motion resonance (MMR) for WASP-43 b and TrES-3 b is estimated to be 1.826 and 1.504 Earth mass, respectively. By quadratic fitting, it is confirmed that WASP-43 b may have a long-term TTV which means an orbital decay. The decay rate is shown to be $\dot{P} =(-0.005248 \pm 0.001714)$~s$\cdot$yr$^{-1}$, and compared with the previous results. Based on this, the lower limit of the stellar tidal quality parameter of WASP-43 is calculated to be $Q'_{\rm *} \ge 1.5\times10^5$, and the remaining lifetimes of the  planets are presented for the different $Q'_{\rm *}$ values of the two systems, correspondingly.}


\maketitle

\section{INTRODUCTION} 

Up to May 2016, about 2600 exoplanets, in which WASP-43 b and TrES-3 b are included, have been discovered with the transit method (http,//exoplanet.eu). When a transiting planet sweeps the surface of its primary star, a part of the light radiated from the star may be transiently obscured. Such an effect manifests a decrease of star flux in observation. By synthesizing the data of transit and radial velocity observations, may be determined the mass and radius of an exoplanet, and furthermore its density, which is a very important parameter for the study of interior structures of exoplanets.

Hellier et al. $^{[1]}$ found that WASP-43 has a transiting companion of planetary size. The mass of this planet is 1.78 Jupiter mass, its revolution period around the star is about 0.81 d, and the orbital semi-major axis is about 0.014 au. Among all the known hot Jupiters it is the nearest one to its primary. Hellier et al. $^{[2]}$ estimated that the number of hot Jupiters with such a short revolution period is less than that of the same kind of objects with the periods “stacking up” around 3$\sim$4 d by two orders of magnitude. In the same way as WASP-19 b$^{[3]}$, WASP-43 b can be taken as an example to study the tidal interaction of a planet with its primary star and the remaining lifetime of the planet. Brownet al. $^{[4]}$ suggested that WASP-19 b accelerates the spin of its primary star, and is in the last stage of spirally falling down to its primary. Barkeret al. $^{[5]}$ pointed out that the expected value of $Q'_{\rm *}$ may depend upon the mass of the troposphere of a star (being the function of $T_{{\rm eff}}$), and therefore upon the spectral type of the star. WASP-43 is a K-type main-sequence star, it is later than WASP-19 in the spectral type (the latter is of the G type). As suggested by Hellier et al.$^{[1]}$, WASP-43 b will be an important case to test the theoretical values.

WASP-43 b has been extensively observed and studied by astronomers. The accuracy of the parameters such as the mass and radius of the planet has been significantly raised by Gillon et al.$^{[6]}$ through their multiple high-intensity observations of primary and secondary eclipses. Afterwards,  Blecic et al.$^{[7]}$ discovered for the first time the decay of its orbital period of
$\dot P = (-0.095 \pm 0.036)$~s$\cdot$yr$^{-1}$, based on the mid-transit times given by Gillon et al.$^{[6]}$ and those provided by the amateurs in the transit observation project TRESCA (TRansiting ExoplanetS and CAndidates, see also http,//var2.astro.cz/EN/tresca)$^{[8]}$. And therefore it is found that the remaining lifetime of WASP-43 b is longer than $5 \times 10^5$ yr, and the lower limit of the stellar tidal quality parameter is $Q'_{\rm *} > 12000$. The orbital decay rate has been recalculated by Murgas et al.$^{[9]}$ to be $\dot P = (-0.015 \pm 0.006)$~s$\cdot$yr$^{-1}$ with the data of 5 times of the GTC (Gran Telescopio Canarias) observation added in.  Chen et al. $^{[10]}$ obtained that $\dot P = (-0.09 \pm 0.04)$~s$\cdot$yr$^{-1}$, by reanalyzing the TRESCA data and the additional multi-band primary eclipse data of GROND (Gamma-Ray Burst Optical/Near-Infrared Detector). Recently, by combining the data of 6 observations made by themselves with the previously available light curve data, Jiang et al.$^{[11]}$ obtained that $\dot P = (-0.029 \pm 0.008)$~s$\cdot$yr$^{-1}$, which shows a slow decay of orbital period, and obtained the order of magnitude of $Q'_{\rm *}$ to be about $10^5$. Based on the results of predecessors and in addition to their own 15 sets of new transit data, Hoyer et al.$^{[12]}$ obtained by fitting that $\dot P = (-0.00002 \pm 0.0066)$~s$\cdot$yr$^{-1}$, so they concluded that the period of WASP-43 b is constant, without orbital decay, and thus determined that $Q'_{\rm *} \ge 10^5$.

The TrES-3 system consists of an adjacent G-type dwarf and a hot Jupiter with its orbital period to be 1.3 d, which is one of the known hot Jupiters relatively close to their primary stars, found by O'Donovan et al $^{[13]}$, and also detected by the SuperWASP Survey of Collier Cameron $^{[14]}$. Later,  Sozzetti et al. $^{[15]}$ improved the measuring accuracy of the system data by the new photometric and spectroscopic observations, redetermined the physical parameters of the system, and studied the transit timing variations (TTV) of TrES-3 b. Having carried out 9 follow-up photometric observations, and combining with the transit data given in the previous literature [15], Gibson et al. $^{[16]}$ gave the period ratio between the potential perturbing planet in the system and  TrES-3 b, as a function of the mass upper-limit of the potential perturbing planet. They found that the accuracy of $O-C$ is high enough to detect an  Earth-mass planet located at the 2:1 mean resonance of TrES-3 b with a circular orbit. Christiansen et al. $^{[17]}$ found a long-term variation in the light curve of TrES-3 by means of 7 transit observations, and suggested that it may be caused by starspots. Similarly, Lee et al. $^{[18]}$ obviated the possibility of obviously periodical TTVs in the system through the data analysis of their own 4 observations in combination with the observations made by predecessors and TRESCA, and suggested that the fluctuation of $O-C$ may be caused by the activity of stellar magnetic field. Turner et al.$^{[19]}$ obtained the similar conclusion by analyzing their 9 optical and near-UV observations. Kundurthy et al. $^{[20]}$ did not find the evidence of TTV with the analysis of 11 transit observations and of the mid-transit times in the relevant literature, and gave the maximum mass of the potential planet which may be located at the 1:2 resonance of TrES-3 b to be 0.66 Earth mass. Jiang et al. $^{[21]}$ displayed the results of 5 transit observations of TrES-3 b, and after analyzing the data of mid-transit times in combination with the light curve data of predecessors, they suggested that there exists probably a periodical term of TTV with a single frequency. Va$\check{\rm n}$ko et al. $^{[22]}$ concluded that during a period of 4 yr there is no possibility to exist a periodical TTV signal with an amplitude greater than 1 min by means of many observations in combination with the data of other authors. Their analysis has ruled out the possibility of existence of a planet greater than Earth mass which is located at the 3:1, 2:1, 5:3 and 3:5, 1:2, 1:3 resonances of TrES-3 b. They found furthermore by means of numerical integration that in the system a planet will be dynamically unstable with its semi-major axis in the range of 0.015$\sim$0.05 au, while in the region beyond 0.05 au a planet will exhibit chaotic behaviors in dynamics, and the energy dissipation will raise with the increases of the initial eccentricity and orbital inclination.

Based on all the work mentioned above, two photometric follow-up observations for the primary eclipse of WASP-43 b and four observations for TrES-3 b are performed using the Xuyi Near-Earth Object Survey Telescope, respectively, and the relevant physical parameters of the two systems are calculated. Combining with the data of other authors, the transit timing variations are calculated and discussed, the lower limits of stellar tidal quality parameter of the primaries are deduced, and the remaining lifetimes of the planets are further simulated. In the second section, our observational results and the photometric data treatment will be demonstrated. In the third section, the fitting of light curves will be described. In the forth section, the orbital periods of WASP-43 b and TrES-3 b will be estimated, and the transit timing variations will be analyzed. In the fifth section, $Q'_{\rm *}$ for the two systems and the corresponding remaining lifetimes of the planets will be discussed. In the final section, a conclusion and the prospects for the future work will be presented.

\section{OBSERVATIONS AND DATA TREATMENT}

The two objects are observed with the Near-Earth Object Survey Telescope at the Xuyi Station of Purple Mountain Observatory, Chinese Academy of Sciences. The telescope is a Schmidt optical telescope with an effective aperture of 1.04 m, focal ratio of $f/1.8$ and haloless effective field of view of 3.14\dg. It is the largest in aperture among the same type telescopes in China, and the fifth all over the world; the seeing at the station is better than 1$''$. Its pointing accuracy in actual observations is 8.16$''$\,(rms); the tracking accuracy is better than 1$''$\,(rms) within 10 min under the CCD guidance. During our observations, the telescope was equipped with a 4K $\times$ 4K CCD detector (now upgraded as a 10K $\times$ 10K CCD detector) with its effective field of view (FOV) of 1.94\dg $\times$ 1.94\dg and angular resolution of 1.705$''$/pixel. Now the telescope is equipped with a standard Bessel photometric system and a Sloan digital survey photometric system. The readout time of the camera is about 43.2 s under the 200\,kHz two-channel readout mode.

The data treatment of observed images is performed with the IRAF (Image Reduction and Analysis Facility) software. At first, the FIT image files obtained from the observation are transformed into the FITS image files (both formats belong to the Flexible Image Transport System) by using the RFITS program in the DATAIO package. Then, a pretreatment of CCD images is implemented with the standard procedure of the CCDPROC program in the CCDRED package, including the background subtraction, flat field correction, and cosmic-ray deletion. A main background is obtained by synthesizing about 10 frames of zero-exposure backgrounds  with the method of median merging; about 10 flat fields are measured with the morning and evening twilight in the selected region with a suitable flux and uniform irradiation, then to have them normalized and merged into a main flat field by means of median merging. The dark current of CCD can be neglected because it is lower than 0.007e$^{-}$/p/s@$-$100~$^{\rm o}$C. The initial time of each exposure is recorded at the head of the FITS image file in Beijing Time (8 h advanced to the Coordinated Universal Time (UTC)), then it is modified into the middle time of exposure, and converted to a Julian date (JD).

The fluxes of objective stars and reference stars are obtained by means of aperture photometry and using the APPHOT package in the IRAF software. Before photometry, in order to eliminate the influence of tracking errors of the telescope, all the images are aligned by using IMALIGN, then the positions of objective stars and their reference stars are determined by using FIND. With this algorithm, the central positions of star images are acquired by fitting the edge distribution in the $x$ and $y$ directions with a Gaussian function. Similarly, the full width at half maximum (FWHM) of point spread function of the star image is fitted with a Gaussian function to show the influence of seeing. When the aperture photometry is performed with the PHOT program, generally we will test many sets of the aperture sizes and sky background rings to choose finally the parameter setting which makes the accuracy of light curve highest; meanwhile, the flux of adjacent dim stars must be excluded in a reasonable scope, so that its influence on the photometry may be neglected. In general, the finally selected aperture size is 2$\sim$3 times of FWHM, and it is shown that the photometric result with such a dynamically selected aperture is better than that with an aperture of fixed radius. After testing multiple reference stars, finally the adjacent non-variable star which is most analogous with the objective star in color and luminosity, and free of the flux overflow and the light pollution of peripheral stars is selected as the reference star.  A differential photometry is implemented by a division of both fluxes to obtain the light curve. In general, we will select multiple reference stars to obtain multiple  light curves, then, among them the one which has the least out-of-transit (OOT) dispersion is selected to make the follow-up fitting and calculation. Taking the straight line fitted with the OOT observational data as a standard, all the observed light curves are finally normalized to the relative flux values.

\subsection{WASP-43 b}
On 24th April 2011 and  7th May 2011, two photometric observations on the event of primary eclipse of WASP-43 b were carried out, respectively. During the first observation, as the sky was clear and favorable to photometric observation, the photometric errors were within 2$\sim$3 mmag; while during the second observation on 7th May, because of the relatively lower elevation angle in the observation and the rather undesirable weather condition, the noise was comparatively large, the photometric errors were within 3$\sim$8 mmag. Table 1 gives the observational information in detail (in which RJD = JD $-$ 2450000/d).

\subsection{TrES-3 b}
On 9th October 2010, 29th January 2011, 25th March 2011 and 11th April 2011, four photometric observations on the event of primary eclipse of TrES-3 b were carried out, respectively. Among all the observations, during the first observation, as the sky was most clear and most favorable to photometric observation, the photometric errors were within 2$\sim$3 mmag; during the second observation, the initial elevation angle was relatively lower, the photometric errors were  within 2$\sim$3 mmag; during the third observation, as the weather condition was relatively undesirable, the photometric errors were within 3$\sim$5 mmag; while at the ingress of the fourth observation, because of the thin cloud cover, the image noise was rather large, and in order to remove its influence the data in this interval were omitted in the light curve fitting, thus the photometric errors  were within 2$\sim$5 mmag. In Table 2, the observational information is given in detail.

\begin{table*}[h!]
\centering
\begin{center}
 \caption {The observation log of WASP-43 b}
  \label{tblog1}
\begin{tabular}{ccccc}
\hline
Date(UT)   & Filter & Interval(RJD)/d & Exposure/s & Number of Images\\
\hline
2011-04-24 &  $r'$  & 5676.064$\sim$5676.162  &    50        &    90   \\
2011-05-07 &  $r'$  & 5689.073$\sim$5689.154  & 80, 100, 120 &    53   \\

\hline
\end{tabular}
\end{center}
\end{table*}


\begin{table*}[h!]
\centering
\begin{center}
\caption{The observation log of TrES-3 b}
  \label{tblog2}
\begin{tabular}{ccccc}
\hline
Date(UT)   & Filter & Interval(RJD) & Exposure/s & Number of Images\\
\hline
2010-10-09 &  $r'$  & 5478.995$\sim$5479.071  &    50        &    71   \\
2011-01-29 &  $r'$  & 5591.320$\sim$5591.417  &  30, 50, 80  &    89   \\
2011-03-25 &  $r'$  & 5646.178$\sim$5646.289  &    50, 80    &    90   \\
2011-04-11 &  $r'$  & 5663.167$\sim$5663.268  &    30, 50    & 102(After Removal) \\

\hline
\end{tabular}
\end{center}
\end{table*}

\section{LIGHT CURVE FITTING}

In general, the white noise can hardly dominate the light curves obtained from the ground-based observations, because of the existence of time-correlated red noises caused by a variety of instrumental and atmospheric effects, including the distortion of light curve caused by the difference of spectral type between the objective and reference stars, the variation of stellar image position caused by the gravity deformation of the instrument, the effect of atmospheric refraction, the tracking error; the variation of stellar image size caused by the variation of seeing, and so on. All these noises must be decorrelated by means of model fitting. In addition, during the primary eclipse, the light of the primary star obscured by a planet in its diverse positions may be influenced by the effect of stellar limb darkening. We adopt the quadratic limb-darkening law derived from an one-dimensional stellar atmosphere model$^{[23-25]}$, namely
\begin{equation}
I_\mu/I_1=1-u_1(1-\mu)-u_2(1-\mu)^2\,,
\label{LD}
\end{equation}
in which $\mu=\cos\theta$ ($\theta$ is the included angle between the observer's line of sight and the normal line of a point on the stellar surface), $I_1$ and $I_{\mu}$ are respectively the brightness seen by the observer at the stellar surface center and at the point where the included angle with the normal line is $\theta$; $u_1$ and $u_2$ are the first-order and second-order limb darkening coefficients, respectively.

The TAP (Transit Analysis Package)$^{[26]}$ is applied to the light curve fitting, and the IDL (Interactive Data Language) is used for this package. The red noises are simulated with the wavelet analysis method on the basis of the transit light curve model of Mandel and Agol $^{[27]}$, then, a Monte-Carlo fitting of Markov chain (MCMC) $^{[28]}$ is implemented with the generated likelihood function as the statistical basis. Meanwhile, the EXOFAST of Eastman et al. $^{[29]}$ is also employed by the TAP, so as to realize the Mandel \& Agol model in IDL.

For each light curve, the parameters to be fitted by the TAP include the planet's revolution period $P$, orbital inclination $i$, the ratio between the orbital semi-major axis and the stellar radius $a/R_{\rm *}$, the ratio of the radius of the planet to that of the star $R_{\rm p}/R_{\rm *}$, the mid-transit time   $T_{\rm mid}$, the first-order and second-order limb-darkening coefficients $u_1$ and $u_2$, the orbital eccentricity $e$, the argument of periastron of the orbit $\omega$, the interception on the vertical axis $F_{\rm int}$ and slope $F_{\rm slope}$ of the atmospheric mass which is used to compensate linearly the global change tendency of atmospheric mass, as well as the non-correlative Gaussian white noise $\sigma_{\rm white}$ and time-correlated red noise $\sigma_{\rm red}$ (which is shown as a function of $1/f^\gamma$, in which $f$ is the frequency, and $\gamma$ is assumed to be 1). Among them, $P$ can be determined only by multiple primary eclipse observations, while $e$ and $\omega$ can not be obtained only by the photometry of primary eclipse, and they are generally determined by the radial velocity method or deduced from a combined calculation of primary eclipse with secondary eclipse. Considered that both WASP-43 b and TrES-3 b are hot Jupiters of short period, even probably situated in synchronous rotation owing to the tidal effect, so $e$ is generally assumed to be 0. Before the MCMC fitting is performed with the TAP, it is necessary to set up the initial values for the above-mentioned parameters; and when the fitting is performed with the TAP after the initial values have been selected, all the above parameters can be selected as: (1) fixed completely; (2) completely free variables; (3) variables in accordance with a Gaussian function, namely the a priori Gaussian variables. Furthermore, all the above parameters may be correlated among multiple light curves, provided they are not fixed completely.

As far as the limb darkening effect is concerned, the quadratic limb darkening model is selected in the fitting process as described afterwards, and the initial values of the coefficients $u_1$ and $u_2$ of limb darkening effect are given by the online tool  EXOFAST $^{[29]}$. This program is based on the  table of coefficients of quadratic limb-darkening effect deduced by Claret et al.$^{[25]}$, and according to the effective temperature $T_{\rm eff}$, surface gravitational acceleration $\lg g$, metal abundance [Fe/H] and  micro-turbulence factor $V_{\rm t}$ of the primary star,  we can obtain the result by a calculation of double-linear interpolation at the SDSS $r'$-band that used in our observations.

\subsection{WASP-43 b}
The main parameters of the primary WASP-43 are selected from Gillonet al. $^{[6]}$. Southworth$^{[30]}$ found that there exists probably a difference of
0.1$\sim$0.2 between the best-fit limb-darkening coefficients and the theoretical values obtained from interpolations, hence, when $u_1$ and $u_2$ are fitted, the theoretical values are taken as the a priori Gaussian medians, and that $\sigma=0.5$ is assumed, so that the full width of Gaussian distribution may cover this difference.  Meanwhile, in order to minimize the potential degeneracy among parameters, the period is selected to be fixed as that given by Table 5 of Gillonet al.$^{[6]}$, and the orbital eccentricity $e$ and the argument of periastron $\omega$ are fixed to be 0. Because of the fixed orbital period, the orbital semi-major axis and eccentricity should not be fitted as fully free parameters. Therefore, when the TAP runs, the initial values of the orbital inclination $i$, the ratio of orbital semi-major axis to stellar radius $a/R_{\rm *}$, and the magnitude of errors $\sigma$ are taken from Table 5 of Gillonet al.$^{[6]}$. While the ratio of planetary radius to stellar radius $R_{\rm p}/R_{\rm *}$ (with its initial value taken from Table 5 of Gillonet al. $^{[6]}$) and the mid-transit time $T_{\rm mid}$, as the main parameters to be obtained from light curve fitting, are assumed to be the fully free parameters to obtain their best-fit values by light curve fitting. Table 3 shows the initial parameter setting in detail.

\begin{table*}[h!]
\begin{center}
 \caption{The initial parameter setting of WASP-43 b}
  \label{tbini1}
\begin{tabular}{ccc}
\hline
 Parameter & Initial value  & Range during the MCMC fitting \\
\hline
 $P$/d & 0.81347753        &  Fixed \\
 $i$/\dg & 82.33             & A priori Gaussian with $\sigma=0.20$, linked with all others\\
 $a$/$R_{\rm *}$         & 4.918  & A priori Gaussian with $\sigma=0.053$, linked with all others \\
 $R_{\rm p}$/$R_{\rm *}$ & 0.15945 & Free, linked with all others\\
 $T_{\rm mid}$(RJD)/d         & Set by eye & Free\\
 $u_1$             & 0.6434    & A priori Gaussian with $\sigma=0.5$, not linked \\
 $u_2$             & 0.1171    & A priori Gaussian with $\sigma=0.5$, not linked \\
 $e$                    & 0.0                   & Fixed \\
 ${\omega}$             & 0.0                   & Fixed \\
\hline
\end{tabular}
\end{center}
\end{table*}

The fitting result obtained by using the TAP is shown in Table 4. All the parameters and their errors are listed in the table. In addition, the light curves obtained from the observations and their best-fit models are illustrated in Fig.1.

The error bars of orbital parameters and of the mid-transit time shown in Table 4 are originated from the TAP calculation. 5 MCMC chains with a length of $10^6$ are calculated by using the TAP, and the final result is obtained by the combination of all the MCMC chains. In the result are recorded the values of 15.9\%, 50.0\% and 84.1\% percentages, in which the value of 50.0\% percentage is the median value which is taken as the optimal value, and the values of other two percentages give the upper and lower limits of the error bar. Such an error estimation has been passed successfully the exam in the paper of Gazak et al. $^{[26]}$, and has succeeded in the tests of a few of other literature (for example, in References [9, 11-12, 21, 31], etc.), hence, our error bars obtained here should be in accordance with the quality of the light curve data, and give a reliable error estimation. It is evident after a comparison that in the calculated error range, all our parameters of the system are in good match with the corresponding ones in the previous literature (for example, References [6, 10-11], etc.)
\begin{table*}[h!]
\begin{center}
\caption{The results of light-curve analysis for WASP-43 b}
\label{tbres1}
\begin{tabular}{lcc}
\hline
 \multirow{2}*{Parameter}
   & \multicolumn{02}{c}{Value}\\
   & WASP-43b\_20110424
   & WASP-43b\_20110507 \\
\hline
           $P$/d      & 0.81347753$^\dagger$ & 0.81347753$^\dagger$\\
         $i$/\dg    & 82.32 $^{+0.11}_{-0.11}$ & 82.32 $^{+0.11}_{-0.11}$\\
          $a/R_{\rm *}$     & 4.933 $^{+0.037}_{-0.037}$ & 4.933 $^{+0.037}_{-0.037}$\\
  $R_{\rm p}/R_{\rm *}$     & 0.1619 $^{+0.0053}_{-0.0052}$ & 0.1619 $^{+0.0053}_{-0.0052}$\\
$T_{\rm mid }$(RJD)/d & 5676.10411 $^{+0.00060}_{-0.00060}$ & 5689.1204 $^{+0.0022}_{-0.0024}$\\
            $u_1$     & 0.647 $^{+0.034}_{-0.035}$ & 0.649 $^{+0.035}_{-0.035}$\\
            $u_2$     & 0.120 $^{+0.035}_{-0.035}$ & 0.121 $^{+0.035}_{-0.035}$\\
              $e$     & 0$^\dagger$ & 0$^\dagger$\\
      ${\omega}$      & 0$^\dagger$ & 0$^\dagger$\\
\hline
\end{tabular}
\end{center}
\begin {minipage}[b]{14cm} {\footnotesize
\hspace*{2.5cm}\hspace*{0.3cm}Note: $^\dagger$Fixed during the MCMC fitting}
\end{minipage}

\end{table*}

\subsection{TrES-3 b}
The main parameters of the primary TrES-3 are selected from Sozzetti et al. $^{[15]}$.
Similarly, the theoretical values of $u_1$ and $u_2$ are taken as the a priori Gaussian medians to make fitting, and it is assumed that $\sigma=0.5$. Meanwhile, according to Table 7 of Sozzetti et al. $^{[15]}$, the value of $P$ is chosen to be fixed, and $e$ and $\omega$ are fixed to be 0, so as to minimize the potential degeneracy among parameters. When the TAP runs, the initial values of $i$ and $a/R_{\rm *}$, and the magnitude of errors $\sigma$ are taken from Table 7 of Sozzettiet al. $^{[15]}$. $R_{\rm p}/R_{\rm *}$ and $T_{\rm mid}$ are assumed as fully free parameters to obtain their best-fit values by light curve fitting.  Table 5 gives the initial parameter setting in detail.

\begin{table*}[h!]
\begin{center}
 \caption {The initial parameter setting of TrES-3 b}
  \label{tbini2}
\begin{tabular}{ccc}
\hline
 Parameter & Initial value  & Range during the MCMC fitting \\
\hline
 $P$/d & 1.30618581        &  Fixed \\
 $i$/\dg & 81.85             & A priori Gaussian with $\sigma=0.16$, linked with others \\
 $a$/$R_{\rm *}$         & 5.926  & A priori Gaussian with $\sigma=0.056$, linked with others  \\
 $R_{\rm p}$/$R_{\rm *}$ & 0.1655 & Free, linked with others\\
 $T_{\rm mid}$(RJD)/d          & Set by eyes & Free\\
 $u_1$             & 0.3643    & A priori Gaussian  with $\sigma=0.5$, not linked \\
 $u_2$             & 0.3178    & A priori Gaussian with $\sigma=0.5$, not linked \\
 $e$                    & 0.0                   & Fixed \\
 ${\omega}$             & 0.0                   & Fixed \\
\hline
\end{tabular}

\end{center}
\end{table*}

The parameters fitted by using the TAP and their errors are shown in Table 6, and the light curves obtained from the observations and their best-fit models are illustrated in Fig.2.

Similarly, 5 MCMC chains with a length of $10^6$ are calculated with the TAP, then the final result and errors are obtained by combining all the chains. It is apparent after a comparison that all the parameters of the system are in good match within the error range with the corresponding ones in the previous literature (for example, References [15, 21-22], etc.)

\section{ORBITAL PERIOD AND TRANSIT TIMING VARIATION}

For the convenience to compare with the relevant parameters in the other literature, the online transformation tool of Eastman et al.$^{[32]}$ is used to convert our mid-transit times that obtained above by fitting from the UTC-based JD into the barycentric Julian date under the barycentric dynamical timescale (BJD$_{{\rm TDB}}$). In order to cover the planet's primary eclipse epochs as wide as possible to calculate more precisely the planet's orbital period $P$ and to analyze the potential TTV in the system, besides our data of mid-transit time $T_{\rm mid}$, we have obtained the $T_{\rm mid}$ data of multiple light curves from the transit data published in the literature, and they are homogenously deduced to BJD$_{{\rm TDB}}$.

Using all the data of $T_{\rm mid}$ obtained above, the new planet's primary eclipse epoch and orbital period $P$ may be acquired by a least $\chi^2$ fitting on the following linear function:

\begin{equation}
 T_{\rm c}(E) = T_0 + PE\,,
\label{TL}
\end{equation}
in which, $T_0$ is the reference time; $E$, the planet's primary eclipse epoch (the primary eclipse epoch in the literature where the planet was discovered is generally defined as $E=0$, other primary eclipse epochs are accordingly defined); $T_{\rm c}(E)$, the mid-transit time calculated under the given epoch $E=0$.

\clearpage
\begin{figure}[tbph]
\centering
{\includegraphics[bb=0 0 425 407,scale=0.85]{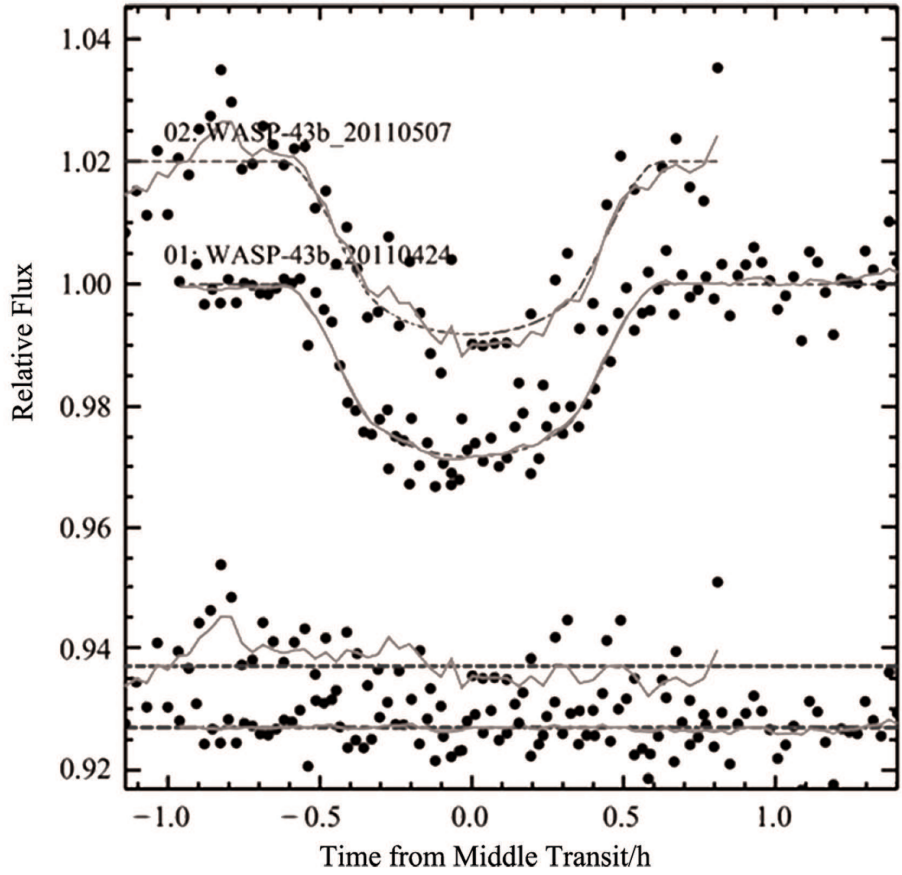}}
\caption{The transit light curves of WASP-43 b and the best-fit models. Each light curve is marked in the figure. The dots are the differential photometry data, the dashed curves are the best-fit models, and the solid curves are the most likely red noise solutions added to the models; the corresponding residuals of the respective fitting are shown at the bottom. Both light curves and residuals are offset on the vertical axis for visual purpose.}
    \end{figure}

\clearpage
\begin{table*}[h!]
\begin{center}
\caption {The results of light-curve analysis for TrES-3 b}
\label{tbres2}
\tabcolsep 1.5mm
\begin{tabular}{lcccc}
\hline
 \multirow{2}*{Parameter}
   & \multicolumn{04}{c}{Value}\\
   & TrES-3b\_20101009
   & TrES-3b\_20110129
   & TrES-3b\_20110325
   & TrES-3b\_20110411 \\
\hline
               $P$/d & 1.30618581$^\dagger$ & 1.30618581$^\dagger$ & 1.30618581$^\dagger$ & 1.30618581$^\dagger$\\
             $i$/\dg & 81.833 $^{+0.077}_{-0.077}$ & 81.833 $^{+0.077}_{-0.077}$ & 81.833 $^{+0.077}_{-0.077}$ & 81.833 $^{+0.077}_{-0.077}$\\
              $a/R_{\rm *}$ & 5.931 $^{+0.028}_{-0.027}$ & 5.931 $^{+0.028}_{-0.027}$ & 5.931 $^{+0.028}_{-0.027}$ & 5.931 $^{+0.028}_{-0.027}$\\
            $R_{\rm p}/R_{\rm *}$ & 0.1634 $^{+0.0080}_{-0.0074}$ & 0.1634 $^{+0.0080}_{-0.0074}$ & 0.1634 $^{+0.0080}_{-0.0074}$ & 0.1634 $^{+0.0080}_{-0.0074}$\\
       $T_{\rm mid}$(RJD)/d & 5479.03441 $^{+0.00088}_{-0.00094}$ & 5591.3682 $^{+0.0015}_{-0.0014}$ & 5646.2272 $^{+0.0020}_{-0.0022}$ & 5663.2029 $^{+0.0027}_{-0.0028}$\\
                $u_1$ & 0.365 $^{+0.025}_{-0.025}$ & 0.366 $^{+0.025}_{-0.025}$ & 0.367 $^{+0.029}_{-0.029}$ & 0.369 $^{+0.029}_{-0.029}$\\
                $u_2$ & 0.318 $^{+0.025}_{-0.025}$ & 0.318 $^{+0.025}_{-0.025}$ & 0.319 $^{+0.029}_{-0.029}$ & 0.320 $^{+0.029}_{-0.029}$\\
                  $e$ & 0$^\dagger$ & 0$^\dagger$ & 0$^\dagger$ & 0$^\dagger$\\
             $\omega$ & 0$^\dagger$ & 0$^\dagger$ & 0$^\dagger$ & 0$^\dagger$\\
\hline
\end{tabular}
\begin {minipage}[b]{14cm} {\footnotesize
\vspace*{1mm}\hspace*{0.3cm}Note: $^\dagger$Fixed during the MCMC fitting}
\end{minipage}
\end{center}
\end{table*}

In order to investigate if there exists a long-term variation of orbital period, a further analysis is carried out. The long-term decay of a close-in planet's orbit implies that there may exist some process of orbital energy dissipation, such as the tidal effect and others $^{[33-34]}$. A simple model under the assumption that the period has a constant variation is given as the following quadratic function:
\begin{equation}
 T_{\rm c}(E) = T_0 + PE + \delta P\times E(E-1)/2\,.
 \label{TQ}
\end{equation}
This model was suggested by Adamset al.$^{[33]}$, in which $\delta P=P\dot{P}$.

The optimal parameters in Eqs.(2,3) are obtained by the least-$\chi^2$ fitting, and the solution is derived by the Levenberg-Marquardt’s least square algorithm which is realized by the MPFIT package of Markwardt$^{[35]}$ based on IDL. In the succeeding comparisons, the optimal model will be chosen according to the Bayes information criterion (BIC), namely,
\begin{equation}
 {\rm BIC} = \chi^2 + k\lg N\,,
 \label{BIC}
\end{equation}
in which, $k$ is the number of free parameters; $N$, the number of data points to be fitted. The finally adopted model is that with the minimum BIC value.

\subsection{WASP-43 b}
The above-mentioned methods and procedures for data treatment and light curve fitting, as well as the software in use are basically as same as those of Jiang et al. $^{[11, 21]}$. In order to ensure the identity of data acquisition so as to raise the accuracy of calculation,  we have selected the primary eclipse epochs and $T_{\rm mid}$ data from Table 5 of Jiang et al. $^{[11]}$ (including the 22 sets of Gillonet al. $^{[6]}$, 1 set of Chen et al. $^{[10]}$, 2 sets of Maciejewskiet al. $^{[36]}$, 1 set of Murgas et al. $^{[9]}$, 5 sets of Ricci et al. $^{[37]}$ and 8 sets of Jiang et al. $^{[11]}$), which are combined with our first set obtained above (the second set is out of use due to its large error caused by the weather and others factors), as well as the 9 sets of Hoyeret al. $^{[12]}$, 6 sets of Stevenson et al. $^{[38]}$  and 1 set of Hellieret al. $^{[1]}$, totally 56 sets of data are fitted, the results are as follows:

For the linear fitting of Eq.(2), the best-fit values are

(1) $T_0 = (5528.8686 \pm 0.00003461)$(BJD$_{TDB}$-2450000);

(2) $P = (0.81347403 \pm 2.657 \times 10^{-8})$ d;

(3) $\chi^2 = 514.21834$ (the degree of freedom is 54), so BIC = 517.71. The standard deviation of $O-C$ is about 44 s.

For the quadratic fitting of Eq.(3), the best-fit values are

(1) $T_0 = (5528.8686 \pm 0.00003508)$(BJD$_{TDB}$-2450000);

(2) $P = (0.81347405 \pm 4.214 \times 10^{-8}$) d;

(3) $\delta P = (-1.353 \times 10^{-10} \pm 4.418 \times 10^{-11}$) d$\cdot$epoch$^{-2}$, so $\dot{P} = \delta P/P = (-0.005248 \pm 0.001714$) s$\cdot$yr$^{-1}$;

(4) $\chi^2 = 511.00842$ (the degree of freedom is 53), so BIC = 516.25. The standard deviation of $O-C$ is about 44 s.

Thus, after a synthetical analysis of our data and the data in the newly published literature, we prefer slightly the quadratic curve model rather than the linear model for the fitting of the mid-transit times ($T_{\rm mid}$) of WASP-43 b, namely, from the analysis on the presently existing data we tend to think that there exists a long-term TTV, or an orbital decay, but the BIC difference between the two is rather small. Our data used for fitting and the resultant $O-C$ after fitting are shown in Table 7, and the distribution of $O-C$ data are illustrated in Fig.3. It is clear from a comparison that within the error range the period $P$ obtained through linear fitting is in match with the previous literature (for example, References [6, 10-11], etc.). As for the result obtained by quadratic fitting, the time span of our data used for fitting is so long that more than 2300 primary eclipse epochs are covered, it is similar to Jiang et al. $^{[11]}$ and Hoyer et al.$^{[12]}$, but longer than Blecic et al. $^{[7]}$, Chen et al. $^{[10]}$ and Murgas et al.$^{[9]}$, which used about 1000 epochs for fitting, by more than two times, hence our fitting result of $\delta P$ is close to those of former ones, but one order of magnitude less than those of latter ones, it corresponds to a decay rate $\dot{P}$ of different order of magnitude. A further discussion on the planetary orbit decay will be given in the next section.

In order to confirm whether there exists a periodic TTV in the system so as to search other objects in it, the discrete Fourier transform program PERIOD04$^{[39]}$ is applied to the $O-C$ of WASP-43 b in the frequency domain within the Nyquist frequency range in the light of Lee et al. $^{[18]}$, and it is found that the highest peak of frequency is 0.002816, and the signal-to-noise ratio of the corresponding amplitude is 2.6. According to the empirical results of observational analysis $^{[40]}$ and of numerical simulation $^{[41]}$, only when the signal-to-noise ratio of amplitude attains at least 4.0 can a good confidence be obtained, but this signal-to-noise ratio has not attained the criterion of confirmation. Therefore, we are inclined to consider that the evidence of periodic TTV signals is not found in the $O-C$ distribution of the WASP-43 system according to the analysis on the data available now.
\begin{figure}[tbph]
\centering
{\includegraphics[bb=0 0 417 544,scale=0.8]{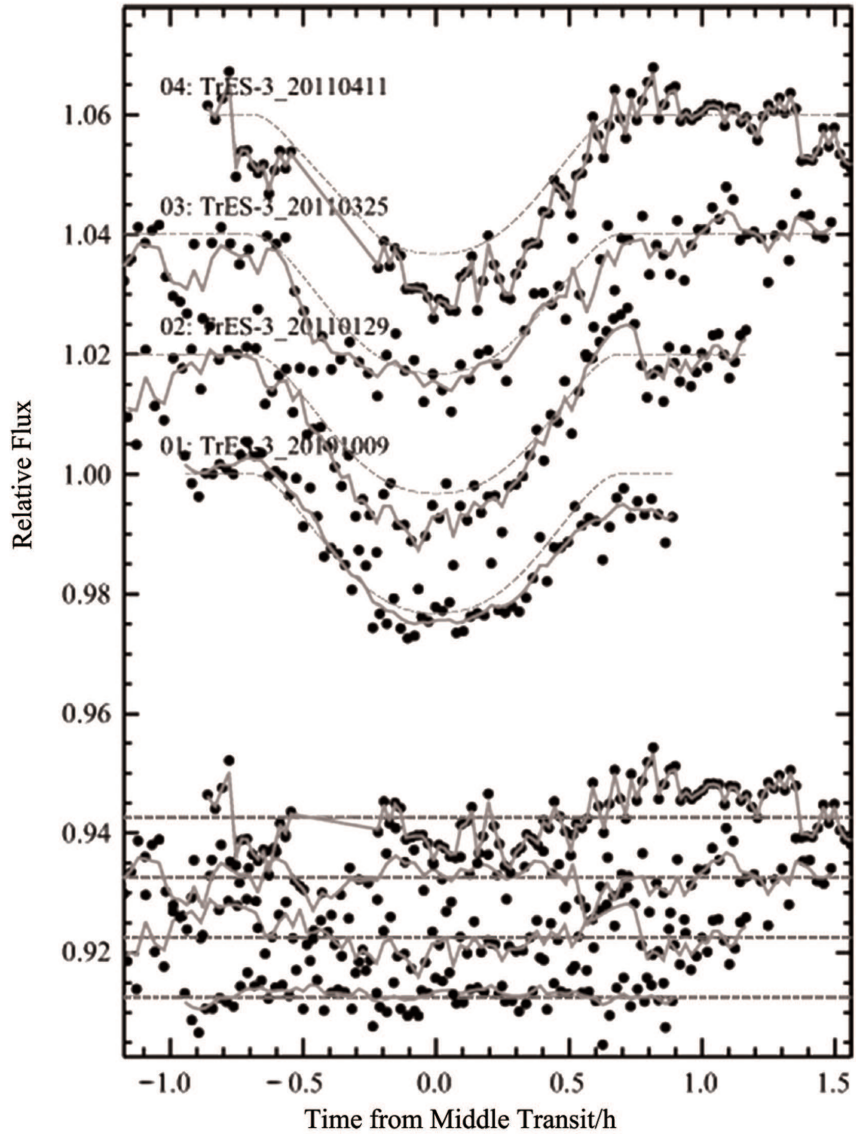}}
\caption{The transit light curves of TrES-3 b and the best-fit models. Each light curve is marked in the figure. The symbols have the same meanings as in Fig.1.}
    \end{figure}

For the case when two planets are located at the $j:(j+1)$ mean motion resonance (MMR), an analytical formula has been given by Agol et al. $^{[42]}$ to estimate an approximate amplitude $\delta t_{\rm max}$ of the transit timing variation caused by the gravitational disturbance as follows,

\begin{equation}
\delta t_{{\rm max}} \approx \frac{P}{4.5j}\frac{m_{{\rm pert}}}{m_{{\rm pert}}+m_{{\rm trans}}}\,,
\label{TTV}
\end{equation}
in which, $m_{{\rm pert}}$ and $m_{{\rm trans}}$ are the mass of a companion and that of the transiting planet, respectively; $P$, the orbital period of the transiting planet. For the WASP-43 system, with the standard deviation of $O-C$ of 44 s, the period $P = 0.81347403$ d as calculated above, and the mass of WASP-43 b to be $m_{\rm trans}=2.034~M_{\rm Jup}$$^{[6]}$, in which $M_{\rm Jup}$ is the Jupiter's mass, it may be obtained that the maximum mass of the potential planet located near the 1:2 resonance of WASP-43 b is $m_{\rm pert} \le 1.826~M_\oplus$, in which $M_\oplus$ is the Earth's mass. The maximum mass of a potential planet located near the resonances of higher orders may be even larger.

\subsection{TrES-3 b}
Similarly, in order to ensure the identity of data acquisition, we have selected the primary eclipse epochs and $T_{\rm mid}$ data from Table 4 of Jiang et al. $^{[21]}$ (including the 8 sets of Sozzettiet al. $^{[15]}$, 9 sets of Gibson et al. $^{[16]}$, 1 set of Colon et al. $^{[43]}$ and 5 sets of Jiang et al. $^{[21]}$), which are combined with our first two sets obtained in the third section (the later two sets are out of use due to their large errors caused by the weather and other factors), as well as the 4 sets of Lee et al. $^{[18]}$, 7 sets of Christiansen et al. $^{[17]}$, 1 set of Sada et al. $^{[44]}$, 12 sets of Va$\check{\rm n}$ko et al. $^{[22]}$, and 11 sets of Kundurthy et al. $^{[20]}$, totally 60 sets of data are used for fitting, and the results are as follows:

For the linear fitting of Eq.((2), the best-fit values are

(1) $T_0 = (4185.9109 \pm 0.00007960)$(BJD$_{{\rm TDB}}-2450000$);

(2) $P = (1.3061866 \pm 1.120 \times 10^{-7})$ d;

(3) $\chi^2 = 149.72572$ (the degree of freedom is 58), so BIC = 153.28. The standard deviation of $O-C$ is about 62 s.

For the quadratic fitting of Eq.(3), the best-fit values are

(1) $T_0 = (4185.9109 \pm 0.00008140)$(BJD$_{{\rm TDB}}-2450000$);

(2) $P = (1.3061866 \pm 1.406 \times 10^{-7})$ d;

(3) $\delta P = (-5.799 \times 10^{-11} \pm 9.565 \times 10^{-11}$) d$\cdot$epoch$^{-2}$, so $\dot{P} = \delta P/P = (-0.001401 \pm 0.002311$) s$\cdot$yr$^{-1}$;

(4) $\chi^2 = 149.78113$ (the degree of freedom is 57), so BIC = 155.12. The standard deviation of $O-C$ is about 62 s.

From the above synthetical analysis of our data and the data newly released in the literature, we prefer the linear model for the fitting result of $T_{\rm mid}$ of TrES-3 b, namely, we tend to think that no evident long-term TTV or orbital decay has been found in the data available now. Table 8 shows our data used for fitting and the resultant $O-C$ of the fitting, and the distribution of $O-C$ data are illustrated in Fig.4. It is clear from a comparison that within the error range the period $P$ obtained through linear fitting is in match with the previous literature (for example References [15, 21-22], etc.).
\vspace*{3mm}
\begin{figure}[tbph]
\centering
{\includegraphics[bb=0 0 360 263]{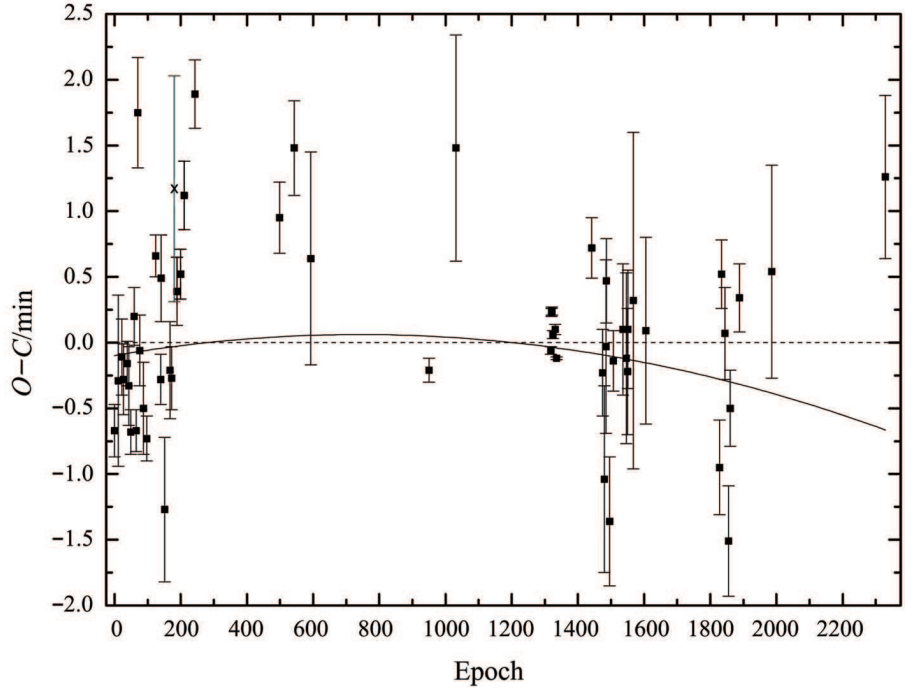}}
\caption{The $O-C$ data and fitting models of WASP-43 b. The squared dots are data from the literature, the cross indicates the data of this paper, the dashed line is the linear $O-C$ fitting model, and the solid line is the quadratic $O-C$ fitting model.}
    \end{figure}

Similarly, the program PERIOD04 is also applied to the $O-C$ data of TrES-3 b in the frequency domain within the Nyquist frequency range to confirm whether  there exists a periodic TTV, and it is found that the highest peak of frequency is 0.04388, which corresponds to the signal-to-noise ratio of amplitude of 3.2, and deviates from the criterion of confirmation that the signal-to-noise ratio attains 4. Therefore, we tend to believe that no significant evidence of periodic TTV signals has been found in the $O-C$ distribution of the TrES-3 system according to the analysis of data now available.
\newpage

\begin{table*}[h!]
\centering
 \caption{The primary eclipse epoch, mid-transit time, and $O-C$ of WASP-43 b}
  \tabcolsep=3pt
  \label{tbo-c1-1}
\begin{tabular}{cccc}
\hline
Epoch  & Source of reference &  $T_{\rm mid}$(${\rm BJD}_{{\rm TDB}}-2450000$)/d & $O-C$/min \\
\hline
11 & [6] & 5537.81659$^{+0.00045}_{-0.00048}$ & $-0.26^{+0.65}_{-0.69}$\\
22 & [6] & 5546.76493$^{+0.00020}_{-0.00021}$ & $-0.08^{+0.29}_{-0.30}$\\
27 & [6] & 5550.83218$^{+0.00019}_{-0.00018}$ & $-0.25^{+0.27}_{-0.26}$\\
38 & [6] & 5559.78048$^{+0.00012}_{-0.00012}$ & $-0.13^{+0.17}_{-0.17}$\\
43 & [6] & 5563.84773$^{+0.00021}_{-0.00020}$ & $-0.30^{+0.30}_{-0.29}$\\
49 & [6] & 5568.72833$^{+0.00012}_{-0.00012}$ & $-0.66^{+0.17}_{-0.17}$\\
59 & [6] & 5576.86368$^{+0.00015}_{-0.00015}$ &  $0.22^{+0.22}_{-0.22}$\\
65 & [6] & 5581.74392$^{+0.00011}_{-0.00011}$ & $-0.65^{+0.16}_{-0.16}$\\
70 & [6] & 5585.81297$^{+0.00029}_{-0.00029}$ &  $1.77^{+0.42}_{-0.42}$\\
76 & [6] & 5590.69256$^{+0.00019}_{-0.00018}$ & $-0.04^{+0.27}_{-0.26}$\\
87 & [6] & 5599.64047$^{+0.00024}_{-0.00024}$ & $-0.47^{+0.35}_{-0.35}$\\
97 & [6] & 5607.77505$^{+0.00012}_{-0.00012}$ & $-0.70^{+0.17}_{-0.17}$\\
124 &[6] & 5629.73981$^{+0.00011}_{-0.00010}$ &  $0.68^{+0.16}_{-0.14}$\\
140 &[6] & 5642.75474$^{+0.00013}_{-0.00013}$ & $-0.26^{+0.19}_{-0.19}$\\
141 &[6] & 5643.56875$^{+0.00023}_{-0.00022}$ &  $0.51^{+0.33}_{-0.32}$\\
152 &[6] & 5652.51574$^{+0.00038}_{-0.00038}$ & $-1.25^{+0.55}_{-0.55}$\\
168 &[6] & 5665.53206$^{+0.00026}_{-0.00026}$ & $-0.20^{+0.37}_{-0.37}$\\
173 &[6] & 5669.59939$^{+0.00017}_{-0.00018}$ & $-0.25^{+0.24}_{-0.26}$\\
189 &[6] & 5682.61543$^{+0.00018}_{-0.00018}$ &  $0.40^{+0.26}_{-0.26}$\\
200 &[6] & 5691.56374$^{+0.00013}_{-0.00013}$ &  $0.54^{+0.19}_{-0.19}$\\
211 &[6] & 5700.51237$^{+0.00018}_{-0.00018}$ &  $1.14^{+0.26}_{-0.26}$\\
243 &[6] & 5726.54407$^{+0.00018}_{-0.00018}$ &  $1.90^{+0.26}_{-0.26}$\\
499 &[10] & 5934.79276$^{+0.00019}_{-0.00019}$ &  $0.95^{+0.27}_{-0.27}$\\
543 &[36] & 5970.58598$^{+0.00025}_{-0.00027}$ &  $1.47^{+0.36}_{-0.39}$\\
593 &[11] & 6011.25910$^{+0.00056}_{-0.00054}$ &  $0.63^{+0.81}_{-0.78}$\\
950 &[9] & 6301.66872$^{+0.00006}_{-0.00005}$ & $-0.24^{+0.09}_{-0.07}$\\
1032&[36] & 6368.37476$^{+0.00060}_{-0.00068}$ &  $1.44^{+0.86}_{-0.98}$\\
1442&[37] & 6701.89857$^{+0.00016}_{-0.00017}$ &  $0.66^{+0.23}_{-0.24}$\\
1475&[11] & 6728.74255$^{+0.00023}_{-0.00024}$ & $-0.30^{+0.33}_{-0.35}$\\
1480&[11] & 6732.80936$^{+0.00049}_{-0.00047}$ & $-1.10^{+0.71}_{-0.68}$\\
1485&[37] & 6736.87743$^{+0.00046}_{-0.00048}$ & $-0.10^{+0.66}_{-0.69}$\\
\hline
\end{tabular}
\end{table*}

\clearpage
\setcounter{table}{6}
\begin{table*}[h!]\tabcolsep=2.73pt
\centering
 \caption{Continued}\smallskip
  \label{tbo-c1-2}
\begin{tabular}{cccc}
\hline
Epoch  & Source of reference &  $T_{\rm mid}$(${\rm BJD}_{{\rm TDB}}-2450000$)/d & $O-C$/min \\
\hline
1486&[37] & 6737.69125$^{+0.00022}_{-0.00022}$ &  $0.40^{+0.32}_{-0.32}$\\
1496&[37] & 6745.82472$^{+0.00034}_{-0.00035}$ & $-1.43^{+0.49}_{-0.50}$\\
1507&[11] & 6754.77378$^{+0.00016}_{-0.00016}$ & $-0.21^{+0.23}_{-0.23}$\\
1550&[37] & 6789.75311$^{+0.00033}_{-0.00033}$ & $-0.29^{+0.48}_{-0.48}$\\
1828&[11] & 7015.89837$^{+0.00025}_{-0.00024}$ & $-1.04^{+0.36}_{-0.35}$\\
1844&[11] & 7028.91466$^{+0.00024}_{-0.00023}$ & $-0.02^{+0.35}_{-0.33}$\\
1855&[11] & 7037.86178$^{+0.00029}_{-0.00028}$ & $-1.60^{+0.42}_{-0.40}$\\
1860&[11] & 7041.92985$^{+0.00020}_{-0.00020}$ & $-0.59^{+0.29}_{-0.29}$\\
1536&[12] & 6778.36469$^{+0.00035}_{-0.00035}$ &  $0.03^{+0.50}_{-0.50}$\\
1546&[12] & 6786.49928$^{+0.00045}_{-0.00045}$ & $-0.19^{+0.65}_{-0.65}$\\
1551&[12] & 6790.56680$^{+0.00031}_{-0.00031}$ &  $0.03^{+0.45}_{-0.45}$\\
1567&[12] & 6803.58254$^{+0.00089}_{-0.00089}$ &  $0.25^{+1.28}_{-1.28}$\\
1605&[12] & 6834.49439$^{+0.00049}_{-0.00049}$ &  $0.01^{+0.71}_{-0.71}$\\
1834&[12] & 7020.78023$^{+0.00018}_{-0.00018}$ &  $0.43^{+0.26}_{-0.26}$\\
1888&[12] & 7064.70770$^{+0.00018}_{-0.00018}$ &  $0.24^{+0.26}_{-0.26}$\\
1986&[12] & 7144.42829$^{+0.00056}_{-0.00056}$ &  $0.44^{+0.81}_{-0.81}$\\
2329&[12] & 7423.45037$^{+0.00043}_{-0.00043}$ &  $1.14^{+0.62}_{-0.62}$\\
1318&[38] & 6601.02729$^{+0.00002}_{-0.00002}$ & $-0.06^{+0.03}_{-0.03}$\\
1320&[38] & 6602.65444$^{+0.00002}_{-0.00002}$ &  $0.23^{+0.03}_{-0.03}$\\
1321&[38] & 6603.46792$^{+0.00002}_{-0.00002}$ &  $0.24^{+0.03}_{-0.03}$\\
1324&[38] & 6605.90822$^{+0.00002}_{-0.00002}$ &  $0.06^{+0.03}_{-0.03}$\\
1332&[38] & 6612.41604$^{+0.00003}_{-0.00003}$ &  $0.10^{+0.04}_{-0.04}$\\
1336&[38] & 6615.66978$^{+0.00001}_{-0.00001}$ & $-0.12^{+0.01}_{-0.01}$\\
0   &[1]  & 5528.86809$^{+0.00014}_{-0.00014}$ & $-0.67^{+0.20}_{-0.20}$\\
181 &This work & 5676.10818$^{+0.00060}_{-0.00060}$ & $1.18^{+0.86}_{-0.86}$\\
197$^\dagger$&This work$^\dagger$ & 5689.1234$^{+0.0022}_{-0.0024}$ $^\dagger$& $0.65^{+3.17}_{-3.46}$ $^\dagger$\\
\hline
\end{tabular}
\begin {minipage}[b]{14cm} {\footnotesize
\vspace*{1mm}\hspace*{0.3cm}Note: $^\dagger$Not used in the TTV fitting due to its large error bar}
\end{minipage}
\end{table*}

For the TrES-3 system, according to Eq.(5), from the standard deviation of $O-C$ of 62 s, the period $P = 1.3061866$ d calculated in the foregoing paragraph, and the mass of TrES-3, namely $m_{\rm trans}=1.910~M_{\rm Jup}$$^{[15]}$, we can obtain that the maximum mass of a potential planet located near the 1:2 resonance of TrES-3 b is $m_{\rm pert} \le 1.504~M_\oplus$. And the potential planets with an even larger maximum mass may exist near the positions of higher-order resonances. Moreover, the TTVs observed in the TrES-3 system also may be caused by the stellar magnetic activity of its primary, such as the spot activity $^{[17-19]}$.

\vspace{5mm}
\begin{figure}[tbph]
\centering
{\includegraphics[bb=0 0 336 245]{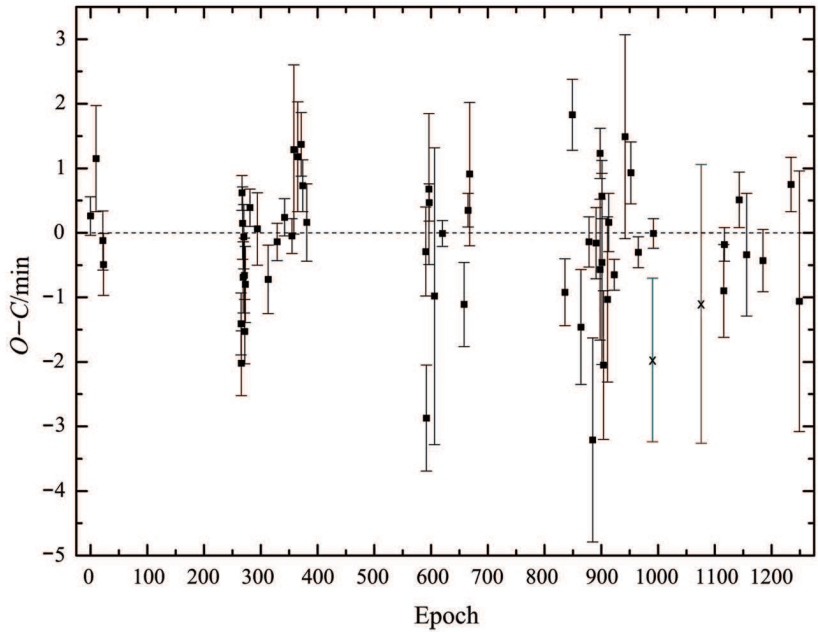}}
\caption{The $O-C$ data and fitting model of TrES-3 b. The squared dots are data from the literature, the crosses are data of this paper, and the dashed line is the linear fitting model of the $O-C$ data.}
    \end{figure}

\section{$Q'_{\rm *}$ AND THE REMAINING LIFETIME OF A PLANET}

According to the tide theory, the close-in transiting planets, such as WASP-43 b and TrES-3 b, may undergo a very strong effect of tidal dissipation (see for example References [45-48], etc.). Such a dissipative effect is caused by the planetary tide and stellar tide. The former plays a dominant role when the planet has a significant eccentricity so as to make the planet's rotation synchronize with its orbital motion and to make its orbit circularized and decayed, while the latter will keep the planet's orbit decay after it has become circular, until the planet spirally falls down to the vicinity of its Roche limit $a_{\rm R}=2.16R_{\rm p}(M_{\rm *}/M_{\rm p})^{1/3}$ and disintegrates $^{[49-50]}$ (in which $R$ is the radius, $M$ is the mass, the subscripts * and p indicate the star and the planet, respectively), because the system can not attain the state of tidal equilibrium due to total angular momentum less than the critical angular momentum (see for example References [1, 34, 51-52], etc.). The present orbits of the two systems that we have observed are very close to a circle (with an eccentricity far less than 0.1), being in the state of stellar tidal evolution. The timescale of tidal evolution with which the present orbit of a planet decays to the vicinity of the Roche limit represents the remaining lifetime of the planet. However, this timescale is not definite, because the stellar tidal dissipation coefficient $Q'_{\rm *}$, which characterizes the tidal effect, can not be determined as it varies within a broad range of several orders of magnitude, it seems absolutely reasonable that $10^6 < Q'_{\rm *} < 10^9$$^{[53]}$.
\begin{table*}[h!]
\centering
 \caption {The primary eclipse epoch, mid-transit time, and $O-C$ of TrES-3 b}
 \tabcolsep=3pt
 \label{tbo-c2-1}
\begin{tabular}{cccc}
\hline
Epoch  & Source of reference &  $T_{\rm mid}$(${\rm BJD}_{{\rm TDB}}-2450000$)/d & $O-C$/min \\
\hline
0   &[15] & 4185.91111$^{+0.00021}_{-0.00021}$ &  $0.26^{+0.30}_{-0.30}$\\
10  &[15] & 4198.97359$^{+0.00057}_{-0.00066}$ &  $1.15^{+0.82}_{-0.95}$\\
22  &[15] & 4214.64695$^{+0.00032}_{-0.00036}$ & $-0.12^{+0.46}_{-0.52}$\\
23  &[15] & 4215.95288$^{+0.00033}_{-0.00031}$ & $-0.49^{+0.48}_{-0.45}$\\
267 &[16] & 4534.66317$^{+0.00019}_{-0.00019}$ &  $0.62^{+0.27}_{-0.27}$\\
268 &[15] & 4535.96903$^{+0.00039}_{-0.00037}$ &  $0.15^{+0.56}_{-0.53}$\\
281 &[15] & 4552.94962$^{+0.00020}_{-0.00022}$ &  $0.39^{+0.29}_{-0.32}$\\
294 &[15] & 4569.92982$^{+0.00039}_{-0.00040}$ &  $0.06^{+0.56}_{-0.58}$\\
313 &[15] & 4594.74682$^{+0.00037}_{-0.00034}$ & $-0.72^{+0.53}_{-0.49}$\\
329 &[16] & 4615.64621$^{+0.00020}_{-0.00021}$ & $-0.14^{+0.29}_{-0.30}$\\
342 &[16] & 4632.62690$^{+0.00020}_{-0.00019}$ &  $0.24^{+0.29}_{-0.27}$\\
355 &[16] & 4649.60712$^{+0.00019}_{-0.00017}$ & $-0.05^{+0.27}_{-0.24}$\\
358 &[16] & 4653.52661$^{+0.00091}_{-0.00092}$ &  $1.29^{+1.31}_{-1.32}$\\
365 &[16] & 4662.66984$^{+0.00059}_{-0.00060}$ &  $1.18^{+0.85}_{-0.86}$\\
371 &[16] & 4670.50709$^{+0.00034}_{-0.00034}$ &  $1.37^{+0.49}_{-0.49}$\\
374 &[16] & 4674.42521$^{+0.00028}_{-0.00028}$ &  $0.73^{+0.40}_{-0.40}$\\
381 &[16] & 4683.56812$^{+0.00042}_{-0.00041}$ &  $0.16^{+0.60}_{-0.59}$\\
665 &[43] & 5054.52523$^{+0.00018}_{-0.00017}$ &  $0.35^{+0.26}_{-0.24}$\\
885 &[21] & 5341.8838$^{+0.0011}_{-0.0010}$    & $-3.21^{+1.58}_{-1.44}$\\
898 &[21] & 5358.86606$^{+0.00076}_{-0.00074}$ & $-0.57^{+1.09}_{-1.07}$\\
901 &[21] & 5362.7847$^{+0.0011}_{-0.00098}$   & $-0.46^{+1.58}_{-1.41}$\\
904 &[21] & 5366.70215$^{+0.00080}_{-0.00077}$ & $-2.05^{+1.15}_{-1.11}$\\
911 &[21] & 5375.84617$^{+0.00089}_{-0.00090}$ & $-1.03^{+1.28}_{-1.30}$\\
592 &[18] & 4959.17138$^{+0.00057}_{-0.00057}$ & $-2.87^{+0.82}_{-0.82}$\\
849 &[18] & 5294.86459$^{+0.00038}_{-0.00038}$ &  $1.83^{+0.55}_{-0.55}$\\
898 &[18] & 5358.86731$^{+0.00027}_{-0.00027}$ &  $1.23^{+0.39}_{-0.39}$\\
\hline
\end{tabular}
\end{table*}
\clearpage
\setcounter{table}{7}
\begin{table*}[h!]\tabcolsep=2.73pt
\centering
 \caption{Continued}\smallskip
  \label{tbo-c2-2}
\begin{tabular}{cccc}
\hline
Epoch  & Source of reference &  $T_{\rm mid}$(${\rm BJD}_{{\rm TDB}}-2450000$)/d & $O-C$/min \\
\hline
901 &[18] & 5362.78541$^{+0.00024}_{-0.00024}$ &  $0.57^{+0.35}_{-0.35}$\\
265 &[17] & 4532.04939$^{+0.00033}_{-0.00033}$ & $-1.41^{+0.48}_{-0.48}$\\
266 &[17] & 4533.35515$^{+0.00035}_{-0.00035}$ & $-2.02^{+0.50}_{-0.50}$\\
269 &[17] & 4537.27463$^{+0.00038}_{-0.00038}$ & $-0.69^{+0.55}_{-0.55}$\\
270 &[17] & 4538.58126$^{+0.00035}_{-0.00035}$ & $-0.06^{+0.50}_{-0.50}$\\
271 &[17] & 4539.88703$^{+0.00040}_{-0.00040}$ & $-0.66^{+0.58}_{-0.58}$\\
272 &[17] & 4541.19261$^{+0.00035}_{-0.00035}$ & $-1.53^{+0.50}_{-0.50}$\\
273 &[17] & 4542.49930$^{+0.00041}_{-0.00041}$ & $-0.80^{+0.59}_{-0.59}$\\
591 &[44] & 4957.86698$^{+0.00048}_{-0.00048}$ & $-0.29^{+0.69}_{-0.69}$\\
596 &[22] & 4964.39859$^{+0.00081}_{-0.00081}$ &  $0.68^{+1.17}_{-1.17}$\\
606 &[22] & 4977.4593$^{+0.0016}_{-0.0016}$    & $-0.98^{+2.30}_{-2.30}$\\
658 &[22] & 5058.44418$^{+0.00045}_{-0.00045}$ & $-1.11^{+0.65}_{-0.65}$\\
668 &[22] & 4569.92982$^{+0.00077}_{-0.00077}$ &  $0.91^{+1.11}_{-1.11}$\\
864 &[22] & 5314.45510$^{+0.00062}_{-0.00062}$ & $-1.46^{+0.89}_{-0.89}$\\
913 &[22] & 5378.45937$^{+0.00031}_{-0.00031}$ &  $0.16^{+0.45}_{-0.45}$\\
923 &[22] & 5391.52067$^{+0.00017}_{-0.00017}$ & $-0.65^{+0.24}_{-0.24}$\\
942 &[22] & 5416.3397$^{+0.0011}_{-0.0011}$    &  $1.49^{+1.58}_{-1.58}$\\
952 &[22] & 5429.40118$^{+0.00033}_{-0.00033}$ &  $0.93^{+0.48}_{-0.48}$\\
965 &[22] & 5446.38075$^{+0.00017}_{-0.00017}$ & $-0.30^{+0.24}_{-0.24}$\\
1116&[22] & 5643.6145$^{+0.0005}_{-0.0005}$    & $-0.90^{+0.72}_{-0.72}$\\
1249&[22] & 45817.3372$^{+0.0014}_{-0.0014}$   & $-1.06^{+2.02}_{-2.02}$\\
597 &[20] & 4965.70463$^{+0.00020}_{-0.00020}$ &  $0.47^{+0.29}_{-0.29}$\\
620 &[20] & 4995.74659$^{+0.00014}_{-0.00014}$ & $-0.01^{+0.20}_{-0.20}$\\
836 &[20] & 5277.88225$^{+0.00036}_{-0.00036}$ & $-0.92^{+0.52}_{-0.52}$\\
878 &[20] & 5332.74263$^{+0.00027}_{-0.00027}$ & $-0.14^{+0.39}_{-0.39}$\\
891 &[20] & 5349.72304$^{+0.00038}_{-0.00038}$ & $-0.16^{+0.55}_{-0.55}$\\
992 &[20] & 5481.64799$^{+0.00016}_{-0.00016}$ & $-0.01^{+0.23}_{-0.23}$\\
1117&[20] & 5644.92119$^{+0.00018}_{-0.00018}$ & $-0.18^{+0.26}_{-0.26}$\\
1143&[20] & 5678.88252$^{+0.00030}_{-0.00030}$ &  $0.51^{+0.43}_{-0.43}$\\
1156&[20] & 5695.86235$^{+0.00066}_{-0.00066}$ & $-0.34^{+0.95}_{-0.95}$\\
1185&[20] & 5733.74170$^{+0.00033}_{-0.00033}$ & $-0.43^{+0.48}_{-0.48}$\\
\hline
\end{tabular}
\end{table*}
\clearpage
\setcounter{table}{7}
\begin{table*}[h!]\tabcolsep=2.73pt
\centering
 \caption{Continued}\smallskip
  \label{tbo-c2-2}
\begin{tabular}{cccc}
\hline
Epoch  & Source of reference &  $T_{\rm mid}$(${\rm BJD}_{{\rm TDB}}-2450000$)/d & $O-C$/min \\
\hline
1234&[20] & 5797.74566$^{+0.00029}_{-0.00029}$ &  $0.75^{+0.42}_{-0.42}$\\
990 &This work & 5479.03425$^{+0.00088}_{-0.00094}$ & $-1.97^{+1.27}_{-1.35}$\\
1076&This work & 5591.3669$^{+0.0015}_{-0.0014}$    & $-1.10^{+2.16}_{-2.02}$\\
1118$^\dagger$&This work$^\dagger$ & 5646.2283$^{+0.0020}_{-0.0022}$ $^\dagger$&  $1.15^{+2.88}_{-3.17}$ $^\dagger$\\
1131$^\dagger$&This work$^\dagger$ & 5663.2048$^{+0.0027}_{-0.0028}$ $^\dagger$& $-4.50^{+3.89}_{-4.03}$ $^\dagger$\\
\hline
\end{tabular}
\begin {minipage}[b]{14cm} {\footnotesize
\hspace*{2.cm}\hspace*{0.3cm}Note: $^\dagger$Not used in the TTV fitting due to its large error bar}
\end{minipage}

\end{table*}

For a transiting exoplanet, to measure the decay rate of its orbital period $\dot{P}$ is a method to directly estimate $Q'_{\rm *}$ (see for example References [12, 52, 54-55], etc.). Concretely speaking, the formula given in the paper of Birkby et.al $^{[55]}$ may be used to calculate the shift of transit time $T_{\rm shift}$ in a certain time interval $T$:


\begin{equation}
T_{\rm shift}=\frac{1}{2}T^2\biggl(\frac{{\rm d}n}{{\rm d}T}\biggr)\biggl(\frac{P}{2\pi}\biggr)\,,
\label{Tsh}
\end{equation}
in which, the rate ${\rm d}n/{\rm d}T$ with which the frequency of orbital mean motion varies with the time is expressed as


\begin{equation}
\frac{{\rm d}n}{{\rm d}T}=\biggl(\frac{{\rm d}n}{{\rm d}a}\biggr)\biggl(\frac{{\rm d}a}{{\rm d}T}\biggr)=-\frac{27}{4}n^2\biggl(\frac{M_{\rm p}}{M_{\rm *}}\biggr)\biggl(\frac{R_{\rm *}}{a}\biggr)^5\biggl(\frac{1}{Q'_{\rm *}}\biggr)\,,
\label{dndt}
\end{equation}
in which, $a$ is the orbital semi-major axis of a planet; $n=2\pi/P$, the frequency of mean orbital motion of the planet. This formula is tenable under the simplified conditions that the relevant tidal frequency is equal to $n$, the orbital eccentricity of the planet is $e=0$, the planet's rotation period is in synchronization with its revolution period, and the rotational angular velocity of the primary star is equal to 0. The minimum value of $Q'_{\rm *}$ may be calculated by substituting $T_{\rm shift}$, which can be obtained by a long-term observation, into the equation.

When a planetary system satisfies the conditions that the planet's orbital eccentricity is small, it rotates synchronously, and the orbital period of the planet is less than the rotation period of the primary (the conditions are basically tenable for an unstable transiting planet), the theoretical formula to calculate the planet's remaining lifetime with a given $Q'_{\rm *}$ is $^{[34]}$:


\begin{equation}
 \tau_{a} \approx  \frac{1}{48}\frac{Q'_{\rm *}}{n}\Big(\frac{a}{R_{\rm *}}\Big)^5\frac{M_{\rm *}}{M_{\rm p}}\,.
 \label{tao}
\end{equation}

This timescale should accord with the evolutionary timescale of the system and the statistical distribution of evolution of known hot Jupiter groups. According to the grouped study on the exoplanets which revolve on circular orbits around their primary stars with a surface convection region, Penev et al. $^{[54]}$ suggested that the statistical probability distribution of remaining lifetimes given by $Q'_{\rm *} \ge 10^7$ has a confidence of 99\%.

Moreover, according to the classical tide theory and under the assumption that the eccentricity of the planet is zero, the mean change of the planet's orbital semi-major axis deduced by Rodr\'{\i}guez et al. $^{[56]}$ may be modified as:


\begin{equation}
 < \frac{{\rm d}a}{{\rm d}T} > = -3na^{-4}\frac{k_{\rm *}}{Q_{\rm *}}\frac{M_{\rm p}}{M_{\rm *}}R_{\rm *}^{5}\,,
 \label{dadt}
 \end{equation}
in which, $k_{\rm *}$ and $Q_{\rm *}$ represent the Love number and tidal quality parameter of a star, respectively. In this paper, the modified stellar tidal quality parameter $Q'_{\rm *}$ $\equiv$ $3Q_{\rm *}/2k_{\rm *}$ is used.

Therefore, based on the derived minimum value of $Q'_{\rm *}$, the remaining lifetimes of WASP-43 b and TrEs-3 b are analyzed respectively according to the different values of $Q'_{\rm *}$ and by means of theoretical calculation and numerical simulation, then a discussion is made on the indefinite parameter $Q'_{\rm *}$.

\subsection{WASP-43 b}
For  WASP-43 b, the lower limit of $Q'_{\rm *}$ given by the previous studies ranges between $10^4$ and $2\times10^5$ $^{[7, 11-12]}$. From Eqs.(6,7), and based on our $\dot{P} \approx -0.0070$ s$\cdot$yr$^{-1}$ (the maximum estimate in consideration of the error influence, see also Hoyeret al. $^{[12]}$) obtained in the foregoing paragraph, we can conclude that when $Q'_{\rm *} \approx 1.5\times10^5$, after 5 yr (about 2300 primary eclipse epochs, corresponding to the time span of the data used for fitting by this paper), $T_{\rm shift} \approx 47$ s, which approaches to the standard deviation of $O-C$ (44 s) obtained by us. The bigger the value of $Q'_{\rm *}$, the smaller reversely the value of $T_{\rm shift}$ caused by the tidal effect, hence our observational result may give that $Q'_{\rm *} \ge 1.5\times10^5$. If the accuracy of $\dot{P}$ can be maintained, and after another 7 yr $T_{\rm shift}$ still keeps in the range of about 40 s, then we can obtain that $Q'_{\rm *} \ge 10^6$.

Assuming that $Q'_{\rm *}$ is respectively $10^5$, $10^6$, $10^7$, $10^8$ and $10^9$, and using the orbital and physical parameters of the system given by Gillonet al. $^{[6]}$,  from Eq.(8) the remaining lifetime of WASP-43 b may be calculated correspondingly to be 0.85 Myr, 8.5 Myr, 85 Myr, 850 Myr and 8.5 Gyr. To make numerical integration on Eq.(9) with the RKF(Runge-Kutta-Fehlberg) algorithm, we can obtain the decay curves of the orbital semi-major axis of WASP-43 b as a function of time as shown in Fig.5. So, the times when WASP-43 is disintegrated by the tidal effect are calculated respectively to be 1.2 Myr, 12 Myr, 120 Myr, 1.2 Gyr and 12 Gyr, corresponding to $Q'_{\rm *}=10^5$, $10^6$, $10^7$, $10^8$, $10^9$, and they are in accordance with the evolution timescales obtained from theoretical calculations. If the conclusion of Penev et al. $^{[54]}$ that $Q'_{\rm *} \ge 10^7$ is adopted, the remaining lifetime of WASP-43 b will be between 100 Myr and 10 Gyr.
\clearpage
\begin{figure}[tbph]
\centering
{\includegraphics[bb=0 0 336 245]{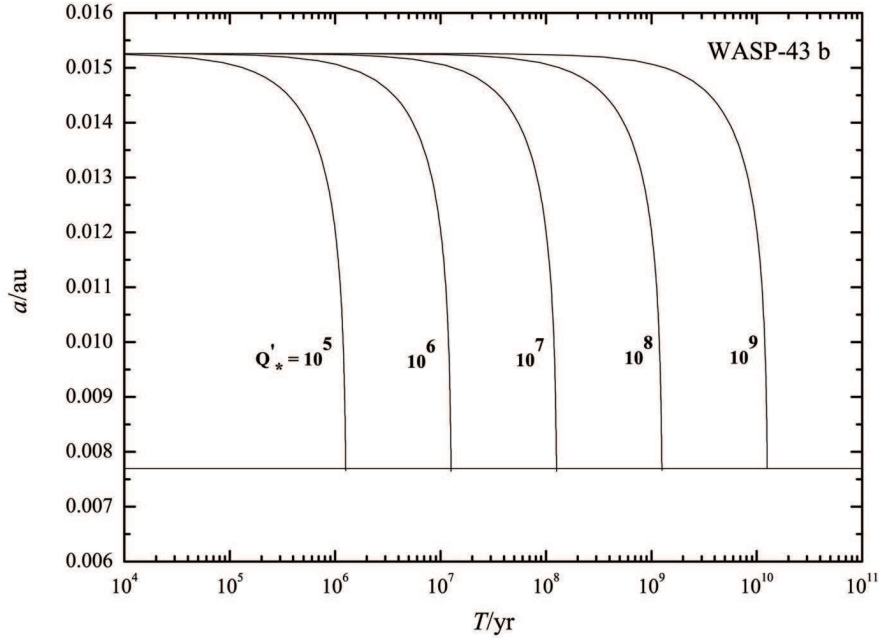}}
\caption{The tidal evolution curves of WASP-43 b for the different values of $Q'_{\rm *}$. The horizontal line below is the Roche limit.}
    \end{figure}

\subsection{TrES-3 b}
For TrES-3 b, based on our $\dot{P} \approx -0.0037$ s$\cdot$yr$^{-1}$ (the maximum estimate after taking the error influence into consideration) obtained in the foregoing paragraph, it is deduced from Eqs.(6,7) that when $Q'_{\rm *} \approx 10^5$, after 4.5 yr (about 1250 primary eclipse epochs, corresponding to the time span of the data used for fitting by this paper), $T_{\rm shift} \approx 10$ s, which is quite different from the standard deviation of $O-C$ (62 s) obtained by us. It may be, therefore, conjectured that there exist probably in the planetary system other factors causing TTV, such as the gravitational disturbance of other objects in the system, the stellar activity and so on as mentioned above.

Assuming that $Q'_{\rm *}$ is respectively $10^5$, $10^6$, $10^7$, $10^8$ and $10^9$, and adopting the orbital and physical parameters of the system given by Sozzettiet al $^{[15]}$,  from Eqs.(6,7) the remaining lifetime of TrES-3 b may be calculated correspondingly to be 4.4 Myr, 44 Myr, 440 Myr, 4.4 Gyr, 44 Gyr. Similar to WASP-43 b, a numerical integration on Eq.(9) is carried out to obtain the tidal evolution curves of TrES-3 b shown in Fig.6. The times when TrES-3 b will be disintegrated by the tidal effect are calculated respectively to be 7.2 Myr, 72 Myr, 720 Myr, 7.2 Gyr and 72 Gyr, which correspond to $Q'_{\rm *}=10^5$, $10^6$, $10^7$, $10^8$, $10^9$, and they are also basically in accordance with the evolution timescales obtained from theoretical calculations. If the conclusion of Penev et al. $^{[54]}$ that $Q'_{\rm *} \ge 10^7$ is adopted, the remaining lifetime of TrES-3 b will be between 700 Myr and 70 Gyr.
\begin{figure}[tbph]
\centering
{\includegraphics[bb=0 0 338 251]{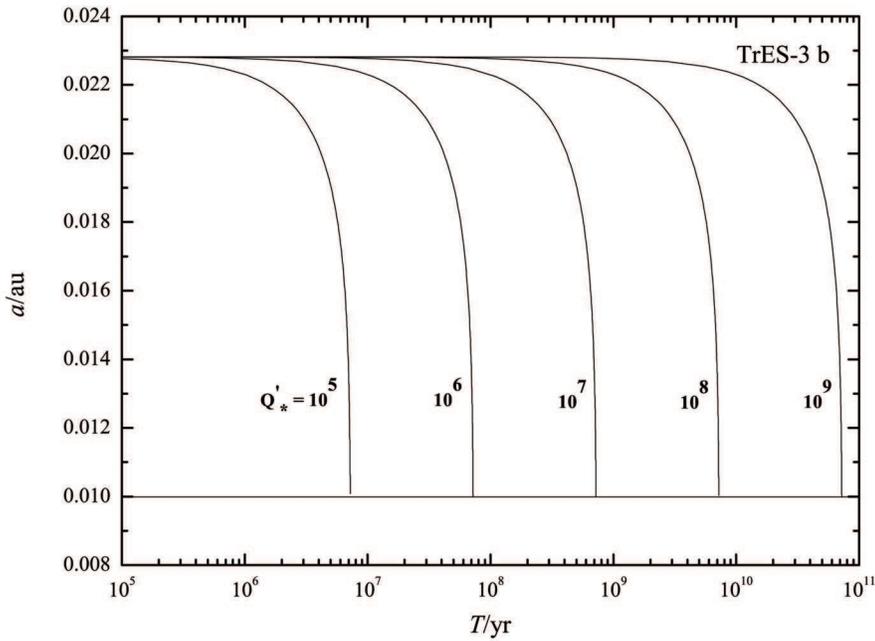}}
\caption{The tidal evolution curves of TrES-3 b for the different values of $Q'_{\rm *}$. The horizontal line below is the Roche limit.}
    \end{figure}

 \section{SUMMARY AND PROSPECT}

Two photometric follow-up transit (primary eclipse) observations on WASP-43 b and four observations on TrES-3 b are performed using the Xuyi Near-Earth Object Survey Telescope of Purple Mountain Observatory. After data treatment and differential photometry the light curves are fitted according to the transit models to obtain the physical parameters of the two systems, which are compared with the results given in the previous literature, it is found that they are in good match within the error range. Combining with the data in a variety of literature, the residuals ($O-C$) of mid-transit times of the two systems are fitted with the linear and quadratic functions, respectively. With the linear fitting, the orbital periods and TTVs of the planets are obtained and discussed. No obvious periodic TTV signal is found in both systems, and the maximum mass of a potential planet located at the 1:2 orbital resonance for WASP-43 b and TrES-3 b is given to be 1.826 and 1.504 masses of Earth, respectively. By quadratic fitting, the long-term TTV, or the orbital decay, is not found for TrES-3 b; while it is revealed that WASP-43 b may have an orbital decay, and the rate of orbital decay for WASP-43 b is shown to be $\dot{P} = (-0.005248 \pm 0.001714)$ s$\cdot$yr$^{-1}$, which is in accordance with the result given in the literature [12], but one order of magnitude less than those in the early literature [7, 9-11]. Based on this, the lower limit of the stellar tidal quality parameter of the WASP-43 system is calculated to be $Q'_{\rm *} \ge 1.5\times10^5$. Finally, the remaining lifetimes of the planets are presented after the respective calculations and simulations with the different values of $Q'_{\rm *}$ for the two systems.
　
The CCD device of the Xuyi Near-Earth Object Survey Telescope used in our observations is of 4K$\times$4K, its angular resolution is 1.705$''$/pixel, and the readout time is 43.2 s, thus the problem of under sampling occurs in both time and space, and it results in a relatively large systematic error and the deficiency of the photometric data quality. Now, the CCD of the Xuyi Near-Earth Object Survey Telescope has been upgraded to 10K$\times$10K pixels with its readout time of about 20 s under some readout modes, so that the sampling rate has been markedly raised, and the photometric data quality is well improved. In addition, to make observation by adopting the defocusing technique can restrain the influence of instrumental effect, and increase the photometric accuracy $^{[57-60]}$. Therefore, we plane to use continually the Near-Earth Object Survey Telescope, and will carry out the follow-up photometric observations of valuable transiting objects with the defocusing technique, as well as the relevant study of TTVs.

As far as transiting exoplanets are concerned, the tidal effect has a significant influence on them. The lower limit of stellar tidal quality parameter $Q'_{\rm *}$ can be determined through the transit observations with long temporal baselines, so as to study the historic and future dynamical evolutions of a planetary system, and therefore to improve our knowledge about the dynamical evolutions of exoplanetary systems. It is of important significance, we shall continue our study of dynamics in this aspect together with the relevant observations.

\acknowledgements{Professor Zhao Hai-bin of Purple Mountain Observatory offered his support and help for our observations of this work, Professor Wang Xiao-bin of Yunnan Observatory gave her advice and aid in our data treatment, we are most grateful to them.}


\begin{thebibliography}{999}

\bibitem{1}
Hellier C., Anderson D. R., Collier Cameron A., et al., A\&A, 2011, 535, L7

\bibitem{2}
Hellier C., Anderson D. R., Collier Cameron A., et al., ApJL, 2011, 730, L31

\bibitem{3}
Hebb L., Collier Cameron A., Triaud A. H. M. J., et al., ApJ, 2010, 708, 224

\bibitem{4}
Brown D. J. A., Collier Cameron A., Hall C., et al., MNRAS, 2011, 415, 605

\bibitem{5}
Barker A. J., Ogilvie G. I., MNRAS, 2009, 395, 2268

\bibitem{6}
Gillon M., Triaud A. H. M. J., Fortney J. J., et al., A\&A, 2012, 542, A4

\bibitem{7}
Blecic J., Harrington J., Madhusudhan N., et al., ApJ, 2014, 781, 116

\bibitem{8}
Poddan\'{y} S., Br\'{a}t L., Pejcha O., NewA, 2010, 15, 297

\bibitem{9}
Murgas F., Pall\'{e} E., Zapatero Osorio M. R., et al., A\&A, 2014, 563, A41

\bibitem{10}
Chen G., van Boekel R., Wang H. C., et al., A\&A, 2014, 563, A40

\bibitem{11}
Jiang I. G., Lai C. Y., Savushkin A., et al., AJ, 2016, 151, 17

\bibitem{12}
Hoyer S., Pall\'{e} E., Dragomir D., et al., AJ, 2016, 151, 137

\bibitem{13}
O'Donovan F. T., Charbonneau D., Bakos G. \'{A}., et al., ApJL, 2007, 663, L37

\bibitem{14}
Collier Cameron A., Wilson D. M., West R. G., et al., MNRAS, 2007, 380, 1230

\bibitem{15}
Sozzetti A., Torres G., Charbonneau D., et al., ApJ, 2009, 691, 1145

\bibitem{16}
Gibson N. P., Pollacco D., Simpson E. K., et al., ApJ, 2009, 700, 1078

\bibitem{17}
Christiansen J. L., Ballard S., Charbonneau D., et al., ApJ, 2011, 726, 94

\bibitem{18}
Lee J. W., Youn J. H., Kim S. L., et al., PASJ, 2011, 63, 301

\bibitem{19}
Turner J. D., Smart B. M., Hardegree-Ullman K. K., et al., MNRAS, 2013, 428, 678

\bibitem{20}
Kundurthy P., Becker A. C., Agol E., et al., ApJ, 2013, 764, 8

\bibitem{21}
Jiang I. G., Yeh L. C., Thakur P., et al., AJ, 2013, 145, 68

\bibitem{22}
Va$\check{\rm n}$ko M., Maciejewski G., Jakub\'{\i}k M., et al., MNRAS, 2013, 432, 944

\bibitem{23}
Claret A., A\&A, 2000, 363, 1081

\bibitem{24}
Claret A., A\&A, 2004, 428, 1001

\bibitem{25}
Claret A., Bloemen S., A\&A, 2011, 529, A75

\bibitem{26}
Gazak J. Z., Johnson J. A., Tonry J., et al., AdAst, 2012, 2012, 697967

\bibitem{27}
Mandel K., Agol E., ApJ, 2002, 580, L171

\bibitem{28}
Carter J. A., Winn J. N., ApJ, 2009, 704, 51

\bibitem{29}
Eastman J., Scott Gaudi B., Agol E., PASP, 2013, 125, 83

\bibitem{30}
Southworth J., MNRAS, 2008, 386, 1644

\bibitem{31}
Sun L. L., Gu S. H., Wang X. B., et al., RAA, 2015, 15, 117

\bibitem{32}
Eastman J., Siverd R., Gaudi B. S., PASP, 2010, 122, 935

\bibitem{33}
Adams E. R., Lopez-Morales M., Elliot J. L., et al., ApJ, 2010, 721, 1829

\bibitem{34}
Levrard B., Winisdoerffer C., Chabrier G., ApJ, 2009, 692, L9

\bibitem{35}
Markwardt C. B., Astronomical Data Analysis Software and Systems XVIII ASP Conference Series, San Francisco: Astronomical Society of the Pacific, 2009, 251

\bibitem{36}
Maciejewski G., Puchalski D., Saral G., et al., IBVS, 2013, 6082, 1

\bibitem{37}
Ricci D., Ram\'{o}n-Fox F. G., Ayala-Loera C., et al., PASP, 2015, 127, 143

\bibitem{38}
Stevenson K. B., Desert J. M., Line M. R., et al., Science, 2014, 346, 838

\bibitem{39}
Lenz P., Breger M., CoAst, 2005, 146, 53

\bibitem{40}
Breger M., Stich J., Garrido R., et al., A\&A, 1993, 271, 482

\bibitem{41}
Kuschnig R., Weiss W. W., Gruber R., et al., A\&A, 1997, 328, 544

\bibitem{42}
Agol E., Steffen J., Sari R., et al., MNRAS, 2005, 359, 567

\bibitem{43}
Colon K. D., Ford E. B., Lee B., et al., MNRAS, 2010, 408, 1494

\bibitem{44}
Sada P. V., Deming D., Jennings D. E., et al., PASP, 2012, 124, 212

\bibitem{45}
Hut P., A\&A, 1981, 99, 126

\bibitem{46}
Eggleton P. P., Kiseleva L. G., Hut P., ApJ, 1998, 499, 853

\bibitem{47}
Dong Y., Ji J. H., SCPMA, 2012, 55, 872

\bibitem{48}
Dong Y., Ji J. H., MNRAS, 2013, 430, 951

\bibitem{49}
Ford E. B., Rasio F. A., ApJ, 2006, 638, L45

\bibitem{50}
Gu P. G., Lin D. N. C., Bodenheimer P. H., ApJ, 2003, 588, 509

\bibitem{51}
Rasio F. A., Tout C. A., Lubow S. H., et al., ApJ, 1996, 470, 1187

\bibitem{52}
Matsumura S., Peale S. J., Rasio F. A., ApJ, 2010, 725, 1995

\bibitem{53}
Jackson B., Greenberg R., Barnes R., ApJ, 2008, 678, 1396

\bibitem{54}
Penev K., Jackson B., Spada F., et al., ApJ, 2012, 751, 96

\bibitem{55}
Birkby J. L., Cappetta M., Cruz P., et al., MNRAS, 2014, 440, 1470

\bibitem{56}
Rodr\'{\i}guez A., Ferraz-Mello S., Michtchenko T. A., et al., MNRAS, 2011, 415, 2349

\bibitem{57}
Southworth J., Hinse T. C., J{\o}rgensen U. G., et al., MNRAS, 2009, 396, 1023

\bibitem{58}
Southworth J., Hinse T. C., Burgdorf M. J., et al., MNRAS, 2009, 399, 287

\bibitem{59}
TAN H. B., WANG X. B., GU S. H., et al., Acta Astronimica Sinica, 2014, 54, 527

\bibitem{60}
Tan H. B., Wang X. B., Gu S. H., et al., ChA\&A, 2014, 38, 307

\end{thebibliography}
\end{document}